\documentclass[journal]{IEEEtran}
\usepackage{cite}
\usepackage{commath}
\usepackage{amsmath,amssymb,amsfonts}
\usepackage{scalefnt}
\usepackage{algorithmic}
\usepackage{graphicx}
\usepackage{textcomp}
\usepackage{nicefrac}
\usepackage{xcolor}
\usepackage{steinmetz}
\usepackage{float}
\usepackage{makecell}
\usepackage{longtable}
\usepackage{subfigure}
\usepackage{multicol}
\usepackage{systeme}
\usepackage{cuted} 
\usepackage{steinmetz}
\usepackage{dblfloatfix}    
\usepackage{tikz}
\usetikzlibrary{shapes.geometric, arrows}
 \tikzstyle{arrow}=[draw, -latex]
\tikzstyle{line}=[draw] 
\usepackage{dblfloatfix}    
\usepackage{tikz}
\usetikzlibrary{shapes.geometric, arrows}
 \tikzstyle{arrow}=[draw, -latex]
\tikzstyle{line}=[draw] 

\tikzstyle{startstop} = [rectangle, rounded corners, minimum width=3cm, minimum height=0.5cm, text centered, draw = black, fill = white]
\tikzstyle{io} = [trapezium, trapezium left angle = 70, trapezium right angle = 110, minimum width = 1cm, minimum height=1cm, text centered, draw = black, fill = white]
\tikzstyle{process} = [rectangle, minimum width=1cm, minimum height=1cm, text centered, draw=black, fill = white]
\tikzstyle{decision} = [diamond, aspect=2, text width=10em, inner sep=-5pt, text centered, draw=black, fill = white]
\tikzstyle{arrows} = [thick, ->, >=stealth]

\ifCLASSINFOpdf
\else
\fi
\begin{document}
%
\title{Analytic estimation of the MMC sub-module capacitor voltage ripple for balanced and unbalanced AC grid conditions}
%
%
%

\author{Daniel Westerman Spier,~\IEEEmembership{Graduate Student Member,~IEEE,} Eduardo Prieto-Araujo,~\IEEEmembership{Member,~IEEE,} Oriol Gomis-Bellmunt,~\IEEEmembership{Senior Member,~IEEE,} and Joaquim L\'opez Mestre}%

\maketitle

\begin{abstract}
In this paper, a mathematical expression to define the maximum and minimum voltage ripples of the modular multilevel converter (MMC) sub-module (SM) capacitors is proposed. Using the arm averaged model of the MMC, the instantaneous power for the upper and lower arms of the converter is obtained, giving the basis to describe the instantaneous energy of the arms. To calculate the SM capacitors peak voltage values, it is required to obtain an analytic expression for the maximum and minimum energy levels. Due to the presence of terms with different magnitudes, frequencies and phases, finding it can be quite challenging. To overcome this issue, the instantaneous arm energy expression is modified using mathematical assumptions in order to join the different components into a single term which can analytically describes the maximum energy point. Then, this expression is used to calculate the peak values of the arm capacitor voltages. By employing the same principles from the arm level, the final analytical expression for the SM capacitors maximum and minimum voltages is found. Simulation results are carried out in order to validate the accuracy of the proposed analysis for different power delivery conditions.

\end{abstract}

\begin{IEEEkeywords}
Capacitor voltage fluctuation, modular multilevel converter (MMC), maximum ripple voltage estimation
\end{IEEEkeywords}

%
\IEEEpeerreviewmaketitle

\section{Introduction}
The modular multilevel converter, firstly proposed by \cite{1304403}, is a well-established technology for HVDC transmission systems \cite{ORIOL,7536194,8013775,8283706}. Among the main benefits compared to other topologies, the MMC  produces high quality output voltage waveforms with low harmonic content, presents reduced transformer $dv/dt$ stress, common DC bus and improved efficiency. Due to these features, the MMC can also be employed in various applications such as medium-voltage motor drives, active filters and microgrids \cite{8013775}.

In microgrids, the MMC can be used in several medium voltage applications \cite{8633435,8286228}. Varying from PV systems \cite{8813638,8472425} to complex interconnection of multiple microgrids \cite{8311251}. Nevertheless, the MMC sub-modules requires larger capacitance compared to other topologies in order to maintain acceptable voltage ripple levels \cite{8610380}. In general, the SM capacitor voltage ripple is assumed to be 10\% of its mean voltage level \cite{6185665}, but this value can vary according to project requirements. Several authors proposed different studies in order to reduce the SM capacitor voltage fluctuation \cite{6642060,7088606,8424186}. However, due to the presence of energy terms with different frequencies, phases and magnitudes, these papers were unable to obtain a concise equation unifying all those elements into a single general expression that can estimate the maximum and minimum energy profiles of the MMC arms.

Having an adequate estimation of the energy ripple can be used during both the design and the operation stages. For instance, during  an  unbalanced  AC  voltage  sag,  the arms of the MMC may present high energy deviations, leading to higher arm capacitor voltages which may exceed the design limitation of the component. Therefore, such expression would be useful to see the impact on the capacitor voltage oscillation when selecting key values of the converter, such as the submodule capacitance. To do so, this paper provides an in-depth mathematical analysis in order to estimate the maximum and minimum submodule capacitor voltages. 

Based on the steady-state analysis of the system given in \cite{IECON}, the different arm variables of the MMC can be obtained. Then, they are used to derive the instantaneous upper and lower arm powers that are later integrated and simplified to achieve the final instantaneous energy expression for the converters arms. Afterwards, mathematical manipulations are performed to combine all the terms with different magnitudes, phases and frequencies into a single equation that can calculate the peak values of the arm capacitor voltage in any grid condition. The same considerations employed in the arm level are extended to the SM capacitors; thus, the final equation for the peak SM capacitor voltages can be achieved.

This paper is divided as: Section \ref{System_Description} describes the converter and its basic equations. Next, in Section \ref{ripple} the instantaneous arm power and energy are derived, giving the basis to obtain the maximum and minimum arm capacitor voltages. In Section \ref{SM_voltage}, the analytical mathematical expression for the SM capacitor peak voltages are demonstrated. Section \ref{results} presents the simulation results to validate the analysis. Finally, the conclusions are given in Section \ref{conclusion}.

\section{System description} \label{System_Description}

The three-phase MMC is a voltage source converter (VSC) consisting of three legs, one per phase, in which each leg has two stacks of $N_{arm}$ half-bridge sub-modules, known as the upper and lower arms. The complexity of the sub-modules can vary according to the application \cite{7656765}. 

\begin{figure}[!h]
\centerline{\includegraphics[width=3.5in,clip]{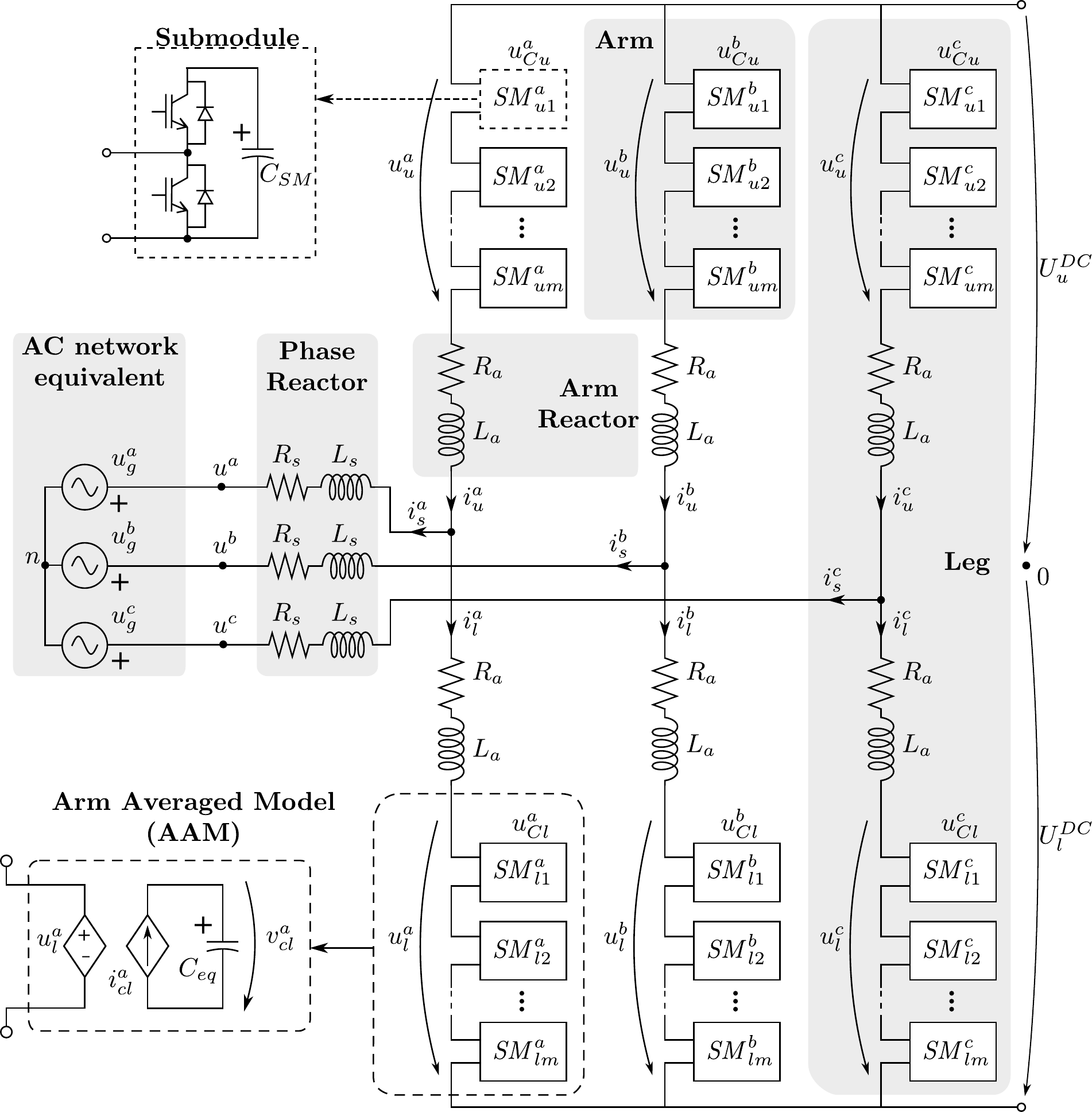}}
\caption{Complete model of the MMC converter.}
\label{fig:MMC}
\end{figure}

In Fig. \ref{fig:MMC}, the converter variables are described as follows: $u_{g}^k$ is the AC network voltage, $u_{u}^k$ and $u_{l}^k$ are the upper and lower arms voltages respectively, $U_{u}^{DC}$ and $U_{l}^{DC}$ are the upper and lower DC grid voltages, $i_{u}^k$ and $i_{l}^k$ are the upper and lower arm currents respectively, $i_{s}^k$ is the AC network current, $R_a$ and $L_a$ are the arm impedances and finally, $R_s$ and $L_s$ are the phase reactor impedances.
~

The mathematical description of the MMC general equations can be obtained per phase $k$ $(\mbox{consider that } k\in \{a,b,c\})$ and is given as,
~
\begin{equation}
U_{u}^{DC} - u_{u}^k - u_{g}^k - u_n = R_{a}i_{u}^k + L_{a}\frac{di_{u}^k}{dt} + R_{s}i_{s}^k + L_{s}\frac{di_{s}^k}{dt}
\label{eq:general_upper}
\end{equation}
~
\begin{equation}
-U_{l}^{DC} + u_{l}^k - u_{g}^k - u_n = -R_{a}i_{l}^k - L_{a}\frac{di_{l}^k}{dt} + R_{s}i_{s}^k + L_{s}\frac{di_{s}^k}{dt}
\label{eq:general_lower}
\end{equation}

\section{Derivation of the arm capacitor voltage ripple} \label{ripple}

The capacitor voltage is highly dependable on the energy flowing through the MMC arms as it can be noted from \eqref{eq:Enormal}. Under balanced AC network conditions, the arms energy are equitable, resulting in similar voltage profiles for the capacitor ripple. However, during an unbalanced AC voltage sag, the arms of the MMC may present high energy deviations, leading to higher arm capacitor voltages which may exceed the design limitation of the component. 

\begin{equation}
E = \dfrac{CU^2}{2}
\label{eq:Enormal}
\end{equation}

Consequently, the capacitor voltages must be kept within certain limits to ensure a proper operation of the MMC. To find the maximum and minimum ripple of the SM capacitor voltage, the derivation procedure is firstly done for the arm capacitors of the MMC and the same principles are then employed to the SM capacitors.

\subsection{Instantaneous MMC arm power}

Firstly, the instantaneous power for the upper and lower arms are firstly derived in (4a) and (4b) considering that the MMC is operated under steady-state conditions and that the power transfer occurs from the DC side to the AC grid. By doing so, the MMC DC currents are flowing in the same direction as the AC ones. 

\begin{subequations}
\begin{equation}
p_u^{k}(t) = u_u^{k}(t)i_u^k(t)
\label{eq:upper_arm_power}
\end{equation}    
\begin{equation}
p_l^{k}(t) = u_l^{k}(t)i_l^k(t) 
\label{eq:lower_arm_power}
\end{equation}
\end{subequations}

By replacing the upper and lower arms voltages and currents by their DC and AC components, respectively, the power equations can be expressed as

\begin{subequations}
\begin{align}
p_u^{k}(t) = \left(U_u^{kDC} - \hat{U}_g^k \cos(\omega t + \theta^k) \right) \cdot \nonumber \\ \cdot \left(I^{kDC} + \dfrac{\hat{I}_s^k \cos(\omega t + \delta^k )}{2} \right) 
\label{eq:upper_arm_power_full}
\end{align}    
\begin{align}
p_l^{k}(t) = \left(U_l^{kDC} + \hat{U}_g^k \cos(\omega t + \theta^k) \right) \cdot \nonumber \\ \cdot \left(I^{kDC} - \dfrac{\hat{I}_s^k \cos(\omega t + \delta^k )}{2}  \right) 
\label{eq:lower_arm_power_full}
\end{align}
\end{subequations}

\noindent with $U_{u,l}^{kDC}$ as the upper and lower arms DC voltages, $I^{kDC}$ as the DC current circulating through the MMC arms, $\delta^k$ as the angle between the AC grid voltages and currents and $\theta^k$ is the phase-shift between the grid voltage phases. By multiplying the terms in \eqref{eq:upper_arm_power_full} and \eqref{eq:lower_arm_power_full}, the upper and lower arms power can be expended as follows

\begin{subequations}
\begin{align}
&p_u^{k}(t) = U_u^{kDC}I^{kDC} + \dfrac{U_u^{kDC}\hat{I}_s^k}{2} \cos(\omega t + \delta^k ) - \\ & - I^{kDC}\hat{U}_g^k \cos(\omega t + \theta^k) - \dfrac{\hat{U}_g^k \hat{I}_s^k}{2}\cos(\omega t + \theta^k) \cos(\omega t + \delta^k )\nonumber 
\label{eq:Uarm_power_comp}
\end{align}    
\begin{align}
&p_l^{k}(t) = U_l^{kDC}I^{kDC} - \dfrac{U_l^{kDC}\hat{I}_s^k}{2} \cos(\omega t + \delta^k ) + \\ & + I^{kDC}\hat{U}_g^k \cos(\omega t + \theta^k) - \dfrac{\hat{U}_g^k\hat{I}_s^k}{2}\cos(\omega t + \theta^k)\cos(\omega t + \delta^k )\nonumber
\label{eq:Larm_power_comp}
\end{align}
\end{subequations}

As it can be observed, the expanded instantaneous power expressions present a term in which two cosines with different phases are multiplied. Using simple trigonometric identities, this term can be simplified as, 

\begin{equation}
\begin{aligned}
&\dfrac{\hat{U}_g^k \hat{I}_s^k}{2}\cos(\omega t + \theta^k) \cos(\omega t + \delta^k ) = \\& = \dfrac{\hat{U}_g^k \hat{I}_s^k}{4} \left[\cos(\theta^k - \delta^k) + \cos(2\omega t + \delta^k + \theta^k)\right]
\end{aligned}  
\label{eq:double_freq_term}
\end{equation}  

Thus, substituting \eqref{eq:double_freq_term} in (6a) and (6b), the power equations can be rewritten as follows

\begin{subequations}
\begin{align}
&p_u^{k}(t) = U_u^{kDC}I^{kDC} + \dfrac{U_u^{kDC}\hat{I}_s^k}{2} \cos(\omega t + \delta^k ) -  \\ & - I^{kDC}\hat{U}_g^k \cos(\omega t + \theta^k) - \dfrac{\hat{U}_g^k \hat{I}_s^k}{4}\cos(\theta^k - \delta^k) - \nonumber \\ 
& \qquad\qquad\quad - \dfrac{\hat{U}_g^k \hat{I}_s^k}{4}\cos(2\omega t + \delta^k + \theta^k) \nonumber
\label{eq:u_arm_power_simplified}
\end{align}    
\begin{align}
&p_l^{k}(t) = U_l^{kDC}I^{kDC} - \dfrac{U_l^{kDC}\hat{I}_s^k}{2} \cos(\omega t + \delta^k ) + \\ & + I^{kDC}\hat{U}_g^k \cos(\omega t + \theta^k) - \dfrac{\hat{U}_g^k \hat{I}_s^k}{4}\cos(\theta^k - \delta^k) - \nonumber \\ 
& \qquad\qquad\quad - \dfrac{\hat{U}_g^k\hat{I}_s^k}{4}\cos(2\omega t + \delta^k + \theta^k) \nonumber
\label{eq:lower_arm_power_simplified}
\end{align}
\end{subequations}

In steady-state conditions, the AC and DC active power exchanged between the grids should be equal, if the semi-conductor losses are neglected. Obviously, this is not feasible with instantaneous values (as AC power is non-constant). However, it is possible to impose an equality between the AC average power (calculated in the phasor domain) and the DC power, as given in \eqref{eq:PDC_PAC}. If this condition is not achieved, the energy in the arms cells would either charge or discharge the arm capacitors and therefore, steady-state conditions would not hold.
~
\begin{equation}
 U_{u,l}^{kDC}I^{kDC}  = \dfrac{\hat{U}_g^k \hat{I}_s^k}{4}\cos(\delta^k - \theta^k)
\label{eq:PDC_PAC}
\end{equation}  

Based on this, the upper and lower arms powers are reduced to their simplest form,
~
\begin{subequations}
\begin{align}
&p_u^{k}(t) = \dfrac{U_u^{kDC}\hat{I}_s^k}{2} \cos(\omega t + \delta^k ) - I^{kDC}\hat{U}_g^k \cos(\omega t + \theta^k) -  \nonumber \\
& \qquad\qquad\qquad - \dfrac{\hat{U}_g^k \hat{I}_s^k}{4}\cos(2\omega t + \delta^k + \theta^k)
\end{align}    
\begin{align}
&p_l^{k}(t) = -\dfrac{U_l^{kDC}\hat{I}_s^k}{2} \cos(\omega t + \delta^k ) + I^{kDC}\hat{U}_g^k \cos(\omega t + \theta^k) - \nonumber \\
& \qquad\qquad\qquad - \dfrac{\hat{U}_g^k\hat{I}_s^k}{4}\cos(2\omega t + \delta^k + \theta^k)
\label{Bosta}
\end{align}
\end{subequations}

\subsection{Arms maximum energy and voltage calculation}

The capacitor voltage presents the same waveform profile as the arm energy. Therefore, to find the maximum and minimum values of the arm capacitor voltage, it is possible to calculate the peak values of the arm energy. The equation for the instantaneous energy flowing through the MMC arms which can be expressed as

\begin{align}
&\qquad\qquad\qquad\qquad  E_u^{k}(t) = \int p_u^{k}(t) dt \nonumber \\
&E_u^{k}(t) = \dfrac{U_u^{kDC}\hat{I}_s^k}{2\omega}\sin(\omega t + \delta^k ) - \dfrac{I^{kDC}\hat{U}_g^k}{\omega} \sin(\omega t + \theta^k) - \nonumber \\
&\qquad \qquad \qquad - \dfrac{\hat{U}_g^k \hat{I}_s^k}{8\omega}\sin(2\omega t + \delta^k + \theta^k) 
\label{upper_energy_full}
\end{align}

\begin{align}
&\qquad\qquad\qquad\qquad  E_l^{k}(t) = \int p_l^{k}(t) dt \nonumber \\
&E_l^{k}(t) = -\dfrac{U_l^{kDC}\hat{I}_s^k}{2\omega}\sin(\omega t + \delta^k ) + \dfrac{I^{kDC}\hat{U}_g^k}{\omega} \sin(\omega t + \theta^k) - \nonumber \\
&\qquad \qquad \qquad - \dfrac{\hat{U}_g^k \hat{I}_s^k}{8\omega}\sin(2\omega t + \delta^k + \theta^k)
\label{lower_energy_full}
\end{align}

\begin{figure}[!h]
\centerline{\subfigure[Pure active power injection]{\includegraphics[width=1.9in,clip]{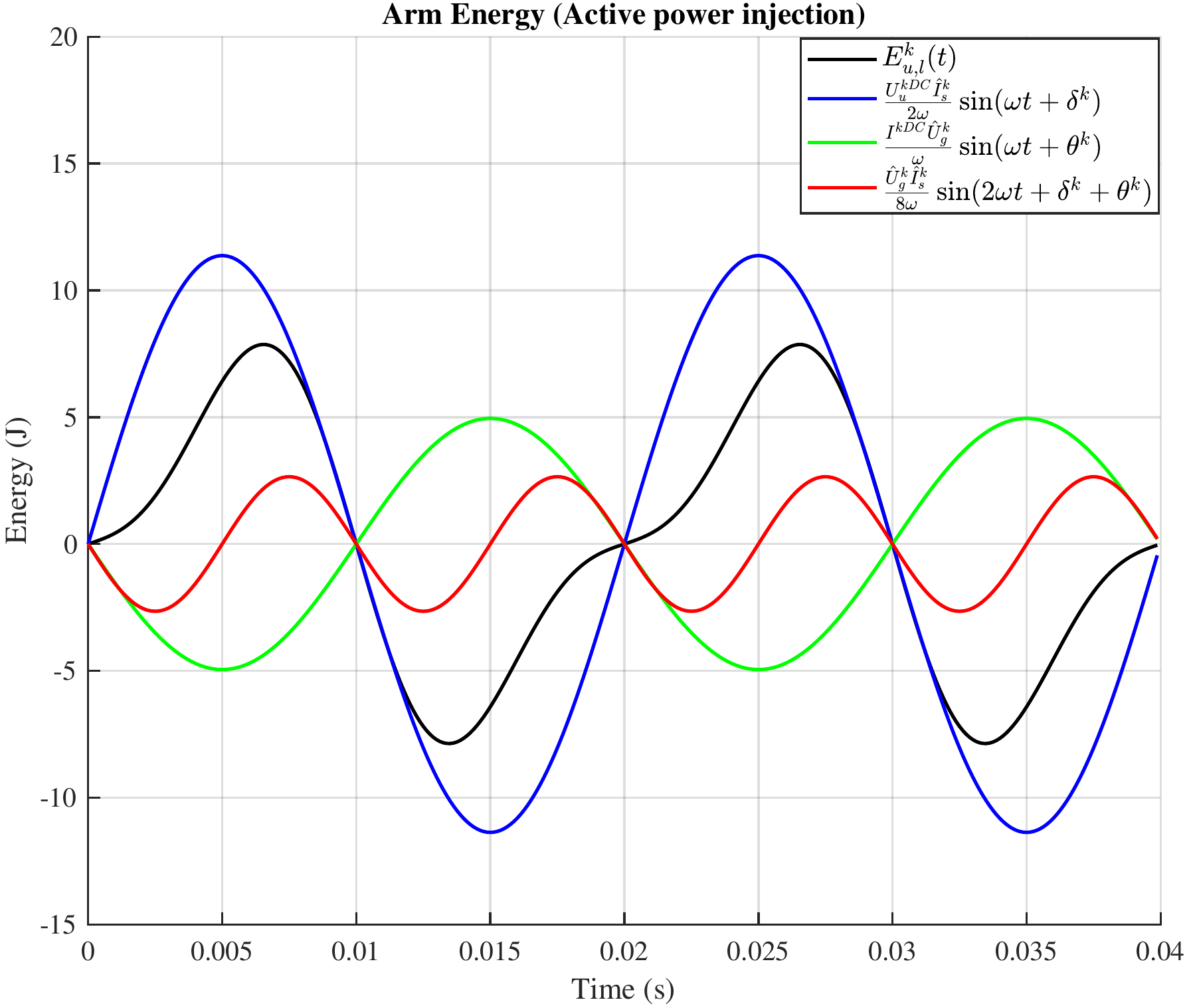}}}
\centering
\subfigure[Pure reactive power injection]{\includegraphics[width=1.9in,clip]{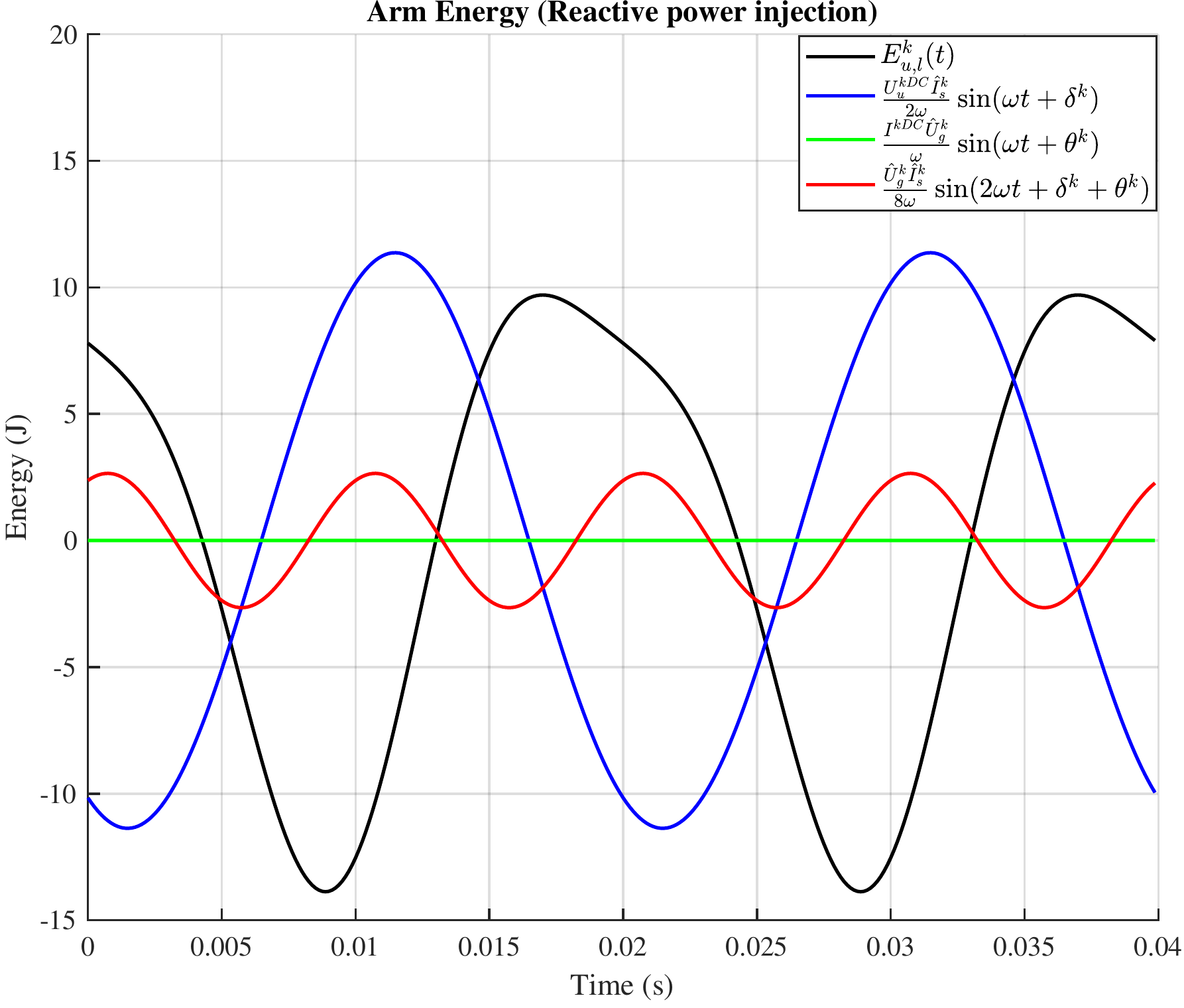}}
\centering
\subfigure[Active and reactive power injection]{\includegraphics[width=1.9in,clip]{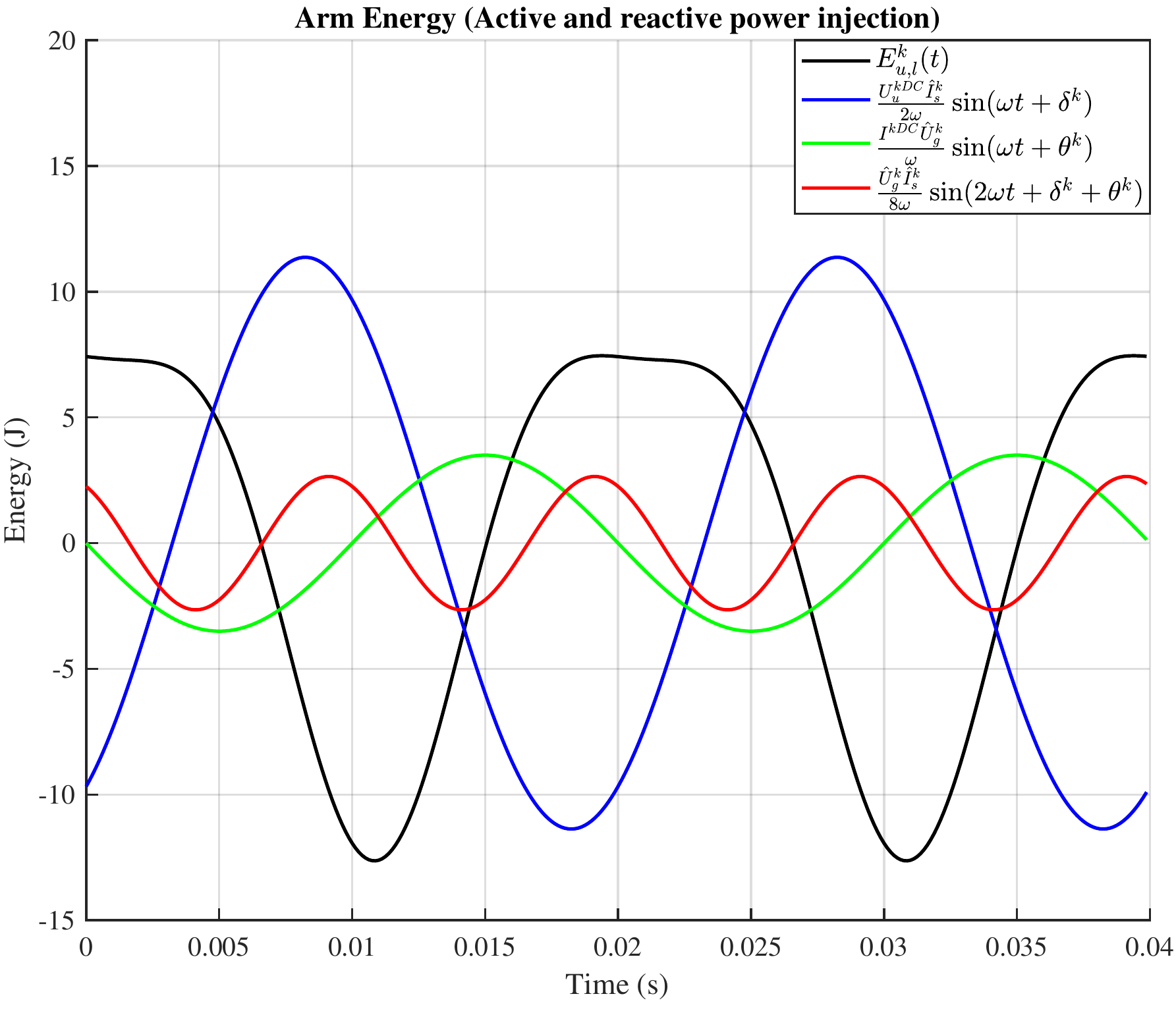}}~
\caption{Energy waveforms for the different energy terms during different power injection conditions.}~
\label{fig:Energy_ripple}
\end{figure}

Even though the expressions for the upper and lower arms instantaneous energy could be obtained through simple mathematical manipulation, finding the maximum value of those can be quite challenging. Such difficult exists due to the presence of terms with different magnitudes, frequencies and phases, as it can be observed in Fig. \ref{fig:Energy_ripple}.
~

The energy expression can be simplified by rewriting the terms with same frequency but with different magnitudes and phases as a single equation by employing the following trigonometric identity,
~
\begin{align}
&\qquad\qquad\qquad A\sin(\omega t + \alpha) + B\sin(\omega t + \beta) = \nonumber\\
&= \underbrace{\sqrt{\left[A\cos(\alpha) + B\cos(\beta) \right]^2 + \left[A\sin(\alpha) + B\sin(\beta) \right]^2}}_\text{X} \cdot \nonumber\\
&\qquad\quad \cdot \sin\left\lbrace\omega t + \underbrace{\tan^{-1}\left[\frac{A\sin(\alpha) + B\sin(\beta)}{A\cos(\alpha) + B\cos(\beta)}\right]}_\text{$\Psi$}\right\rbrace
\label{eq:sine_sum}
\end{align}

Observing \eqref{eq:sine_sum}, it is clear that the resultant expression represents a vector with maximum magnitude equals to $X$ with a phase of $\Psi$. Therefore, the maximum ripple energy value for the terms at the fundamental frequency can be found by replacing the elements from \eqref{upper_energy_full} and \eqref{lower_energy_full} in $X$, such as
%
%


\begin{equation}
E_{umax_f}^k= \left(
\begin{aligned}
\left[\dfrac{U_u^{kDC}\hat{I}_s^k}{2\omega}\cos(\delta^k) - \dfrac{I^{kDC}\hat{U}_g^k}{\omega}\cos(\theta^k) \right]^2 + \\ 
+ \left[\dfrac{U_u^{kDC}\hat{I}_s^k}{2\omega}\sin(\delta^k) - \dfrac{I^{kDC}\hat{U}_g^k}{\omega}\sin(\theta^k) \right]^2
\end{aligned}
\right)^{1/2}
\label{Eu_f_max}
\end{equation}

\begin{equation}
E_{lmax_f}^k= \left(
\begin{aligned}
\left[-\dfrac{U_l^{kDC}\hat{I}_s^k}{2\omega}\cos(\delta^k) + \dfrac{I^{kDC}\hat{U}_g^k}{\omega}\cos(\theta^k) \right]^2 + \\ 
+ \left[-\dfrac{U_l^{kDC}\hat{I}_s^k}{2\omega}\sin(\delta^k) + \dfrac{I^{kDC}\hat{U}_g^k}{\omega}\sin(\theta^k) \right]^2
\end{aligned}
\right)^{1/2}
\label{El_f_max}
\end{equation}

Therefore the upper and lowers instantaneous energies can be described as, 
~
\begin{equation}
E_u^{k}(t) = E_{umax_f}^k \sin(\omega t + \Psi_u^k )  +  E_{max_{2f}}^k\sin(2\omega t + \delta^k + \theta^k) 
\label{upper_energy_full_simp}
\end{equation}
~
\begin{equation}
E_l^{k}(t) = E_{lmax_f}^k \sin(\omega t + \Psi_l^k )  + E_{max_{2f}}^k\sin(2\omega t + \delta^k + \theta^k) 
\label{lower_energy_full_simp}
\end{equation}

\noindent where, $\Psi_u^k$ and $\Psi_l^k$ are the upper and lower phase shifts for phase $k$ obtained using \eqref{eq:sine_sum} and $E_{max_{2f}}^k = \left|-\hat{U}_g^k \hat{I}_s^k/8\omega\right|$.

Although the maximum value of the arms energy at the fundamental frequency was obtained, there is still a need to combine them with the second-order components to achieve a general equation that can estimate the maximum energy ripple of the upper and lower arms. Equations \eqref{upper_energy_full_simp} and \eqref{lower_energy_full_simp} will be maximized when $\sin (\omega t + \Psi_{u,l}^k )= \sin (2 \omega t + \delta^k + \theta^k)=1$; thus,
~
\begin{equation}
E_{u,lmax}^{kAC} = E_{u,lmax_{f}}^k + E_{max_{2f}}^k
\label{eq:E_max_1}
\end{equation}

To comply with the conditions in \eqref{eq:E_max_1}, the angles for the first and second order terms must meet the following condition, 
~
\begin{subequations}
\begin{align}
\centering
\omega t + \Psi_{u,l}^k &= \pi/2 + n_1 \pi, \text{for } n_1 \in \mathbb{Z},  \\
2 \omega t + \delta^k + \theta^k &= \pi/2 + n_2 \pi, \text{for } n_2 \in \mathbb{Z} \\
\pi/2 + n_1 \pi -\Psi_{u,l}^k &=  \left(  \pi/2 + n_2 \pi - (\delta^k + \theta^k) \right) /2 \\
\pi + 2 n_1 \pi -2 \Psi_{u,l}^k &=   \pi/2 + n_2 \pi - (\delta^k + \theta^k) \\
\pi/2 + (2 n_1-n_2) \pi &= 2 \Psi_{u,l}^k-( \delta^k + \theta^k) \\
  2 \Psi_{u,l}^k - (\delta^k + \theta^k)  &= \pi/2 + n \pi 
\end{align}
\label{eq:miticas}
\end{subequations}

\begin{figure}[!h]
\centerline{\includegraphics[width=3.5in,clip]{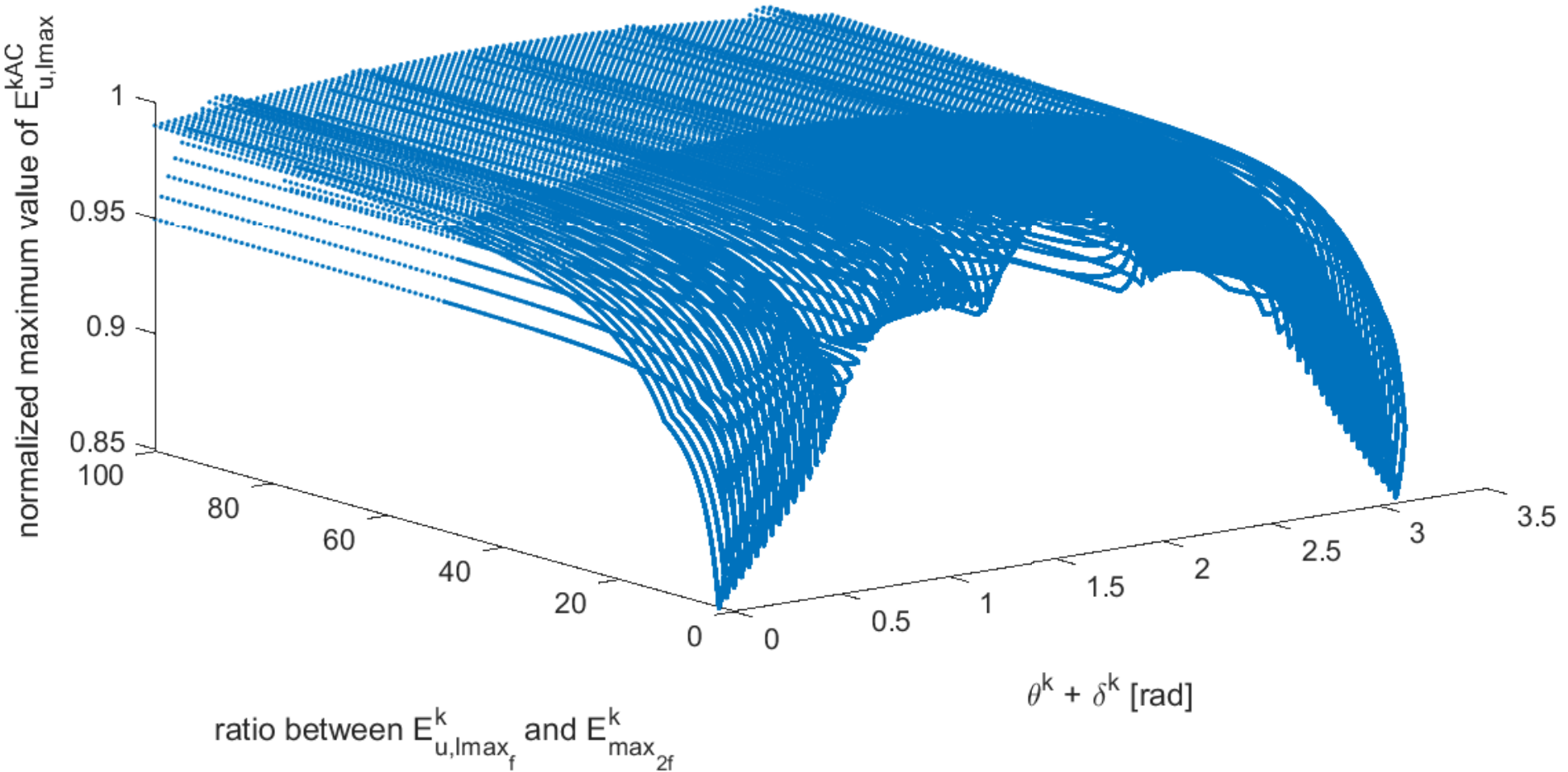}}
\caption{Energy profile for different angle and magnitude values.}
\label{fig:Cap_mitico}
\end{figure}

However, when such condition is not held, the maximum energy level calculated with \eqref{eq:E_max_1} presents a value deviation. This error is affected by the magnitudes of the energy terms, as well as, the phase displacement between them. In Fig. \ref{fig:Cap_mitico}, it can be observed that the maximum energy value ($z$ axis) and, consequently, the minimum error (see \eqref{eq:E_max_1}) happens when the angles for the second order term sum $90^o$ or for high energy ratios, which is in accordance with \eqref{eq:miticas}. Whereas, the highest energy deviation occurs when there is no phase shift between $E_{u,lmax_{f}}^k$ and $E_{max_{2f}}^k$ and small energy ratio, having a value equals to 15\%.

As aforementioned, the analysis considers that the MMC is operated in steady-state conditions and, as a result, the capacitors can be considered fully charged. For this reason, they already present an average energy level for the upper and lower arms \eqref{eq:E_ref}. Moreover, this energy reference must be taken into account when the arm capacitor voltage ripple is calculated.

\begin{equation}
E_{u,l_{ref}}^k = \dfrac{C_{SM}}{2N_{arm}}U_{Cu,l_{ref}}^{k^2}
\label{eq:E_ref}
\end{equation} 

\noindent where $U_{Cu,l_{ref}}^k$ is the arm voltage reference, which is the value of the DC link voltage and $C_{SM}$ is the sub-module capacitance. 

Therefore, the maximum and minimum values for the upper and lower arms energy can be mathematically described as

\begin{equation}
E_{u,l_{max,min}}^k = \underbrace{E_{u,l_{ref}}^k}_\text{DC term} \pm   \underbrace{E_{u,l_{max}}^{kAC}}_\text{peak of the AC  part}
\label{eq:Eul_max_min}
\end{equation}
%

By replacing the energy DC term $E_{u,l_{ref}}^k$ with the maximum/minimum energy ones $E_{u,l_{max,min}}^k$, \eqref{eq:Eul_max_min} can be rearranged and the final expression for the peak voltage levels of the MMC arm capacitor can be calculated as,


\begin{equation}
U_{u,l_{max}}^k = \sqrt{\dfrac{2E_{u,l_{max}}^kN_{arm}}{C_{SM}}}
\label{eq:U_arm_max}
\end{equation}

\begin{equation}
U_{u,l_{min}}^k = \sqrt{\dfrac{2E_{u,l_{min}}^kN_{arm}}{C_{SM}}}
\label{eq:U_arm_min}
\end{equation}

\section{SM capacitor maximum and minimum voltages} \label{SM_voltage}

Until now, the derivation procedure was done in the arm level. However, as aforementioned, the main goal is to obtain a mathematical expression for the maximum and minimum voltages of the individual SM capacitors of the MMC. Based on the arm expressions, the same principles are used for the SM capacitors and the final general expression for their peak voltages can be calculated. Again, it is considered that the MMC is operated under steady-state conditions to ensure that all the SM capacitors are fully charged.

The first step is to modify the upper and lower arms energy expressions given in \eqref{upper_energy_full} and \eqref{lower_energy_full} into equations that can describe the energy profile in the SM level. This can be easily done by dividing those expression by the number of SM per arm, as 
~
\begin{align}
&\qquad\qquad\qquad\qquad  E_u^{k}(t) = \int p_u^{k}(t) dt \nonumber \\
&E_{SMu}^{k}(t) = \dfrac{U_u^{kDC}\hat{I}_s^k}{2\omega N_{arm}}\sin(\omega t + \delta^k ) - \dfrac{I^{kDC}\hat{U}_g^k}{\omega N_{arm}} \sin(\omega t + \theta^k) - \nonumber \\
&\qquad \qquad \qquad - \dfrac{\hat{U}_g^k \hat{I}_s^k}{8\omega N_{arm}}\sin(2\omega t + \delta^k + \theta^k) 
\label{eq:SM_upper_energy_full}
\end{align}
~
\begin{align}
&\qquad\qquad\qquad\qquad  E_l^{k}(t) = \int p_l^{k}(t) dt \nonumber \\
&E_{SMl}^{k}(t) = -\dfrac{U_l^{kDC}\hat{I}_s^k}{2\omega N_{arm}}\sin(\omega t + \delta^k ) + \dfrac{I^{kDC}\hat{U}_g^k}{\omega N_{arm}} \sin(\omega t + \theta^k) - \nonumber \\
&\qquad \qquad \qquad - \dfrac{\hat{U}_g^k \hat{I}_s^k}{8\omega N_{arm}}\sin(2\omega t + \delta^k + \theta^k)
\label{eq:SM_lower_energy_full}
\end{align}

Then, employing similar procedures used to obtain the maximum energy for the upper and lower arms \eqref{eq:E_max_1} the maximum SM energy levels can be expressed as follows 
~
\begin{subequations}
\begin{align}
\centering
E_{SMu_{max}}^{kAC} &= E_{SM_{umax_{f}}}^k +  E_{SM_{umax_{2f}}}^k \\
E_{SMu_{max}}^{kAC} &=\dfrac{E_{umax}^{kAC}}{N_{arm}} 
\end{align}
\label{eq:SM_upper_E_max}
\end{subequations}
~
\begin{subequations}
\begin{align}
E_{SMl_{max}}^{kAC} &= E_{SM_{lmax_{f}}}^k +  E_{SM_{lmax_{2f}}}^k \\
E_{SMl_{max}}^{kAC} &= \dfrac{E_{lmax}^{kAC}}{N_{arm}} 
\end{align}
\label{eq:SM_lower_E_max}
\end{subequations}

The DC energy levels for SM capacitors can be described modifying \eqref{eq:E_ref} as
~
\begin{equation}
E_{SM{u,l_{ref}}}^k = \dfrac{C_{SM}}{2}U_{SM}^2
\label{eq:E_SM_ref}
\end{equation} 

\noindent where $U_{SM}$ is the average SM capacitor voltage.

Analogously to the maximum and minimum arm energies \eqref{eq:Eul_max_min}, the peak energy values for the SM capacitors can be calculated as, 

\begin{equation}
E_{SMu,l_{max,min}}^k = E_{SM{u,l_{ref}}}^k \pm E_{SMu,l_{max}}^{kAC}
\end{equation}

Finally, the analytic expressions to described the maximum and minimum voltages for the SM capacitors are obtained removing the number of SM in \eqref{eq:U_arm_max} and \eqref{eq:U_arm_min}, resulting in

\begin{equation}
U_{Cu,l_{max}}^k = \sqrt{\dfrac{2E_{SMu,l_{max}}^k}{C_{SM}}}
\label{eq:UC_max}
\end{equation}

\begin{equation}
U_{Cu,l_{min}}^k = \sqrt{\dfrac{2E_{SMu,l_{min}}^k}{C_{SM}}}
\label{eq:UC_min}
\end{equation}

\section{Results} \label{results}
In this section, the proposed analytical expressions to calculate the peak values of the SM capacitor voltages are compared with the maximum and minimum levels obtained using the full energy expressions given in \eqref{eq:SM_upper_energy_full} and \eqref{eq:SM_lower_energy_full}. Firstly, the MMC phase variables were determined through the steady-state analysis developed in \cite{IECON} using the parameters presented in Table \ref{tab:param_v2}, which corresponds to an application where the MMC was used as the interface between the main AC distributed grid with different local AC and DC networks. Based on those values, the energy for the SM capacitors were obtained and used to calculate the peak values of the SM capacitor voltages.

In order to validate the proposed mathematical analysis, the converter was considered to be operated in all four different quadrants. Moreover, the different power transfer conditions were analyzed based on balanced AC and DC grid conditions and the displayed quantities are related to phase $a$.

\begin{table}[ht]
\caption{System parameters}\renewcommand\arraystretch{1} 
\begin{tabular}[c]{lcccl}
\hline\hline
\textbf{Parameter}                   & \textbf{Symbol} & \textbf{Value} & \textbf{Units}     \\ \hline
Rated power                          & $S$               & 10           & kVA                \\
AC-side voltage                      & $U_g$               & 400            & V rms ph-ph \\
DC link voltage                    & $U^{DC}$             & $\pm$350           & V  \\
Phase reactor impedance              & $Z_s$          & j0.24        & $\Omega$                 \\
Arm reactor impedance                & $Z_a$             & j1.57     & $\Omega$                                \\
Converter modules per arm            & $N_{arm}$            & 8            & -                  \\
Average module voltage               & $U_{SM}$         & 87.5            & V                \\
Sub-module capacitance               & $C_{SM}$         & 1            & mF                \\
\hline\hline
\end{tabular}
\label{tab:param_v2}
\end{table}

In Table \ref{tab:Results_1}, the maximum and minimum SM capacitor voltages are shown when the values are obtained using the full energy expression and using the proposed peak energy equation. Whereas, in Table \ref{tab:Results_2} the values for the fundamental and second order frequencies energy terms, as well as, their respective phases are given for the proposed analysis. Finally, the errors for the different power delivery scenarios are displayed in Table \ref{tab:Results_3}. 

Observing Table \ref{tab:Results_2}, it can be noted that the value of the second-order energy term is practically constant for different scenarios, since the RMS levels for the grid current and voltage are the same for all the cases. On the other hand, the other quantities changed depending on the power transfer condition as the DC terms and phases have to change in order to absorb or inject reactive power. 

The negative signal in the errors for the maximum SM capacitor voltage indicates that the values obtained by the proposed analysis were smaller than the actual level, which can be considered a safety factor during the design stage.  Meanwhile, the positive signal in the errors for the minimum voltages implies that the estimated magnitudes are also lower than the real ones. Contrarily to the maximum voltage case, the prediction of lower minimum voltages is undesirable since it can affect the normal operation of the MMC depending to the circumstances. Therefore, in cases where the converter is required to inject reactive power to the AC grid, the addition of a factor of safety to reduce the error, is recommended, and it can be assessed using Fig. \ref{fig:Cap_mitico}.

The values obtained in Tables \ref{tab:Results_3} are in agreement with the proposed analysis. The two scenarios where the errors are equal to zero are the same as predicted in \eqref{eq:miticas}. For the other power delivery conditions, the error profile respects the ones observed in Fig. \ref{fig:Cap_mitico}.

\begin{table}[!t]
\caption{Results (Part I)}
\centering
\begin{tabular}{*{5}{c}}
\hline\hline
\multicolumn{1}{c} {\textbf{Power references}} & \multicolumn{2}{c}{\textbf{Full equation} [V]} & \multicolumn{2}{c}{\textbf{Proposed method} [V]} \\ 
\cline{2-5}
\multicolumn{1}{c}{[kW]} & \multicolumn{1}{c}{Maximum} & \multicolumn{1}{c}{Minimum} & \multicolumn{1}{c}{Maximum} & \multicolumn{1}{c}{Minimum} \\ \hline 

P = 10, Q = 0 & 98.410 & 75.823 & 99.644 & 73.372 \\ 
P = 7.07, Q = 7.07 & 97.415 & 68.788 & 103.25 & 68.205 \\
P = 0, Q = 10 & 99.175 & 64.424 & 105.65 & 64.424\\
P = -7.07, Q = 7.07& 97.415 & 68.788 & 103.250 & 68.205 \\ 
P = -10, Q = 0& 98.410 & 75.823& 99.6445 & 73.3732  \\
P = -7.07, Q = -7.07 & 103.24 & 76.47 & 102.917 & 68.7067\\
P = 0, Q = -10 & 105.65 & 74.005 & 105.65 & 64.424\\
P = 7.07, Q = -7.07 & 103.24 & 76.47 & 102.917& 68.707 \\
\hline\hline
\end{tabular}
\label{tab:Results_1}
\end{table}

\begin{table}[!t]
\caption{Results (Part II)}
\centering
\begin{tabular}{*{4}{c}}
\hline\hline
 \multicolumn{1}{c}{$E_{SM_{umax_{f}}}^a$ [J]} & \multicolumn{1}{c}{$E_{SM_{umax_{2f}}}^a$ [J]} &\multicolumn{1}{c}{$\Psi^a$ [rad]} &\multicolumn{1}{c}{($\theta^a+\delta^a$) [rad]}\\
\hline 
0.8047 & 0.3316 & -0.0490 & 0.0637 \\ 
1.1707 & 0.3315& -1.0644 & -0.7423 \\
1.4213 & 0.3316 & 1.5708 & -1.5708\\
1.1707 & 0.3315 & 1.0644 & -2.3993 \\ 
0.8047 & 0.3316 & 0.0490 & -3.2053  \\
1.1363 & 0.3315 & -1.0475 & 2.3090\\
1.4213 & 0.3316 & 1.5708 & 1.5708\\
1.1363 & 0.3315 & 1.0475& 0.8326\\
\hline\hline
\end{tabular}
\label{tab:Results_2}
\end{table}

\begin{table}[!t]
\caption{Error between the values from the full equation and the suggested analysis}
\centering
\begin{tabular}{*{3}{c}}
\hline\hline
\multicolumn{1}{c} {\textbf{Power references}} &\multicolumn{2}{c}{\textbf{Error [\%]}} \\ \cline{2-3}
\multicolumn{1}{c}{[kW]} & \multicolumn{1}{c}{Maximum} & \multicolumn{1}{c}{Minimum} 
\\ 
\hline
P = 10, Q = 0 & -1.253 & 3.231\\ 
P = 7.07, Q = 7.07 & -5.990 & 0.848  \\
P = 0, Q = 10 & -6.529 & 0\\
P = -7.07, Q = 7.07& -5.99 & 0.848 \\ 
P = -10, Q = 0& -1.253 & 3.231 \\
P = -7.07, Q = -7.07 & 0.3125 & 10.15 \\
P = 0, Q = -10 & 0 & 12.946 \\
P = 7.07, Q = -7.07 &0.313 & 10.152 \\
\hline\hline
\end{tabular}
\label{tab:Results_3}
\end{table}

\section{Conclusions} \label{conclusion}

In this paper, the MMC sub-module capacitors voltage was analyzed and a mathematical estimation of their maximum and minimum voltages was proposed. By considering the arm averaged model, the instantaneous power and energy arm equations were derived. Using mathematical assumptions, the terms with different magnitudes, frequencies and phases were transformed into a single equation which could describe the peak values of the arm capacitor energy ripple for different scenarios.

In steady-state conditions, the arm capacitors can be considered fully charged, consequently, they present a DC energy term that was added with the energy ripple. Then, the summed expression was used to calculate the maximum and minimum voltage ripples of the arm capacitor. Based on the process used for arm capacitors, the SM capacitor instantaneous energy was derived. Furthermore, similar procedures were employed to simplify the SM capacitor energy, allowing the final expression for the SM capacitor peak voltages to be calculated. 

Simulation results were carried out to validate the proposed analysis for different power delivery conditions. The results have showed that the SM capacitor voltage ripples were in agreement with the profiles established during the derivation procedures. Finally, the analysis developed by this paper can be employed in future works in order to improve the current reference calculation techniques which may include capacitor voltage limitation issues in any grid voltage condition.

\section{Acknowledgments}
This project has received funding from the European Union's Horizon 2020 research and innovation programme under the Marie Sklodowska-Curie grant agreement no. 765585. This document reflects only the author’s views; the European Commission is not responsible for any use that may be made of the information it contains. This work was partially supported by the Spanish Ministry of Science, Innovation and Universities under the Project RTI2018-095429-B-I00. This work was co-financed by the European Regional Development Fund. E. Prieto is lecturer of the Serra H\'unter Programme.

\ifCLASSOPTIONcaptionsoff
  \newpage
\fi

\bibliographystyle{IEEEtran}
\bibliography{MYBIBI_OPTIMIZATION.bib}
%




\end{document}